\documentclass[10pt]{iopart}   
\usepackage{epsfig}
\usepackage{graphics}

\newcommand{\sitei}{\vec{i}}
\newcommand{\sitej}{\vec{j}}
\newcommand{\sitek}{\vec{k}}
\newcommand{\I}{{\rm i}}
\newcommand{\diff}{{\rm d}}
\newcommand{\sgn}{{\rm sgn}}

\begin{document}

\title[DDMRG for the Mott--Hubbard insulator in high
  dimensions]{Dynamical Density-Matrix Renormalization Group\\ 
for the Mott--Hubbard insulator in high dimensions}

\author{Satoshi Nishimoto\dag, Florian Gebhard\dag, Eric Jeckelmann\ddag}

\address{\dag\ Fachbereich Physik, Philipps-Universit\"at Marburg,
D-35032 Marburg, Germany}

\address{\ddag\ Institut f\"ur Physik, KOMET 337, 
Johannes Gutenberg-Universit\"at Mainz, 55099 Mainz, Germany}

\ead{Florian.Gebhard@Physik.Uni-Marburg.DE}

\begin{abstract}
We study the Hubbard model at half band-filling on a Bethe lattice 
with infinite coordination number in the paramagnetic insulating phase
at zero temperature.
We use the dynamical mean-field theory (DMFT) mapping to a single-impurity
Anderson model with a bath whose properties
have to be determined self-consistently. 
For a controlled and systematic
implementation of the self-consistency scheme we use the 
fixed-energy (FE) approach to the DMFT. In FE-DMFT
the onset and the width of the Hubbard bands
are adjusted self-consistently but the energies of the bath levels
are kept fixed relatively to both band edges
during the calculation of self-consistent 
hybridization strengths between impurity and bath sites.
Using the dynamical density-matrix renormalization group method
(DDMRG) we calculate the density of states 
with a resolution ranging from 3\% of the bare bandwidth 
$W=4t$ 
at high energies to 0.5\% in the vicinity of the gap.
The DDMRG resolution and accuracy for the density of states
and the gap is superior to those obtained with other
numerical methods in previous DMFT investigations.
We find that the Mott gap closes at a critical coupling 
$U_{\rm c}/t=4.45 \pm 0.05$.
At $U=4.5t$, we observe prominent shoulders near the onset
of the Hubbard bands. They are
the remainders of the quasi-particle resonance in the metallic phase
which appears to split when the gap opens at $U_{\rm c}$.
\end{abstract}



\submitto{\JPCM}


\section{Introduction}

In the limit of high dimensions~\cite{dinfty}, 
models for correlated lattice electrons
can be mapped onto effective single-impurity 
problems~\cite{BrandtMielsch,Jarrell}.
In some cases, the exact solution for a many-particle Hamiltonian
has been found, e.g., for the Falicov-Kimball 
model~\cite{BrandtMielsch,Freericksreview},
and for other problems a few exact results have been obtained; 
for reviews, see~\cite{RMP,Buch}. 
Despite its increasing popularity~\cite{RMP,Vollirev,Phystoday}
it must be kept in mind that 
the dynamical mean-field theory (DMFT)
still poses a difficult many-body problem:
the effective single-impurity model 
must be solved self-consistently for
the one-particle Green function at all frequencies.
Consequently, reliable numerical or analytical `impurity solvers' 
must be developed and the self-consistency scheme must be
implemented in a controlled way.

The Hubbard model at half band-filling 
provides an ideal test case for the dynamical mean-field theory. 
It describes $s$-electrons
with a purely local interaction of strength~$U$ and
electron transfer matrix elements~$-t/\sqrt{Z}$ 
between~$Z\to\infty$ neighboring sites
on a lattice.
On the one hand, 
the model contains an interesting quantum phase transition
between the paramagnetic metal and the paramagnetic (Mott--Hubbard)
insulator~\cite{Mottbuch} at a finite coupling. On the other hand,
for a Bethe lattice with a semi-elliptic bare density of states
of width~$W=4t$,
perturbation theory to fourth order in $U/W$~\cite{mottmetal}
and to second order in~$W/U$~\cite{mottinsu} have been carried out
at zero temperature,
against which approximate analytical and numerical techniques
can be tested. In this way, merits and limitations of
analytical methods (Hubbard-III approximation~\cite{HubbardIII},
iterated perturbation theory~\cite{IPT}, local moment 
approach~\cite{LMA}) and numerical techniques
[exact diagonalization schemes (ED)~\cite{ED-CK,ED-RK}, 
numerical renormalization
group (NRG)~\cite{NRG}] have been revealed, together with the
difficulties in implementing the self-consistency scheme in
numerical approaches. 

The latter problem results from the fact
that numerical approaches work with a finite number of sites
to represent the continuous bath coupled to the impurity site.
Thus the energy resolution is necessarily limited by 
finite-size effects.
Moreover, it is not clear a priori how one can define a self-consistency
condition for the discretized impurity problem such
that the self-consistent solution 
is approached in a smooth and controlled way
in the limit of an infinite number of bath sites.
[In other approaches to the impurity problem,
such as quantum Monte Carlo (QMC)
simulations, the bath is not discretized
but the imaginary time has to be discretized, which leads
to similar problems.]
In a previous
work~\cite{mottmetal,mottinsu} this problem has been solved
by the `fixed-energy' approach to the dynamical mean-field theory (FE-DMFT):
(i)~a frequency interval~$I$ is split
into subintervals $I_{\ell}$ of equal length, whose
mid-points $\epsilon_{\ell}$ give the energies of the bath sites,
and the density of states is put to zero outside this interval~$I$; 
(ii)~the hybridization strengths between impurity and bath sites
is determined self-consistently for these fixed energies~$\epsilon_{\ell}$.
Within the fixed-energy approach to dynamical mean-field theory,
the resolution of the frequency interval~$I$ improves systematically 
with system size~$n_s$, and an extrapolation
$n_s\to\infty$ becomes meaningful. As has been shown in
Refs.~\cite{mottmetal,mottinsu}, the FE-DMFT combined
with exact diagonalization [FE-DMFT(ED)] with $n_s\leq 15$ provides
a reasonable description of the metallic phase
for $U\leq 0.4 W$ and of the Mott--Hubbard insulator for $U/W \geq 1.2$.

With exact diagonalization 
finite-size effects are prominent
in the interesting region of the metal-insulator transition, i.e.,
for $U\approx W$.
Consequently, numerical approaches are required which overcome
the limitation of the exact diagonalization technique. 
The dynamical density-matrix renormalization group method
(DDMRG)~\cite{Eric}
treats large systems (here with up to $n_s=161$ sites)
very accurately. It is an extension of the standard
density-matrix renormalization group (DMRG)~\cite{White,DMRGbook}
to the calculation of dynamical correlation functions.
For the computation of a continuous spectrum,
DDMRG is more accurate than previous generalizations 
of the DMRG to dynamical
quantities such as the Lanczos-DMRG~\cite{Hallberg},
or the correction-vector DMRG~\cite{SteveundCo}.
The DDMRG has been applied successfully to the single-impurity Anderson
model~\cite{EricSatoshi} and to DMFT calculations
for the metallic phase of the
Hubbard model~\cite{mottmetal}.

In this work, we present results for
the Mott--Hubbard insulator on a Bethe lattice with $Z\to\infty$
neighbors obtained with FE-DMFT combined with DDMRG [FE-DMFT(DDMRG)]. 
In Section~\ref{model}
we specify the Hubbard model,
the effective single-impurity Hamiltonian, 
the corresponding one-particle Green functions, and the
self-consistency condition. We also recall the results
from perturbation theory in $1/U$.
In Section~\ref{method} we summarize important aspects
of the fixed-energy approach to the dynamical mean-field theory,
and the DDMRG impurity solver.
Details can be found in Refs.~\cite{mottmetal,mottinsu,EricSatoshi}.
In Section~\ref{results} we display the density of states,
the gap for single-particle excitations, the ground-state energy and the
average double occupancy as a function of 
the interaction strength $U$ in the
Mott insulating phase found for $U> U_{\rm c}= 4.45 (\pm 0.05) t$. 
A short summary and conclusions close our
presentation. A sum rule for the ground-state energy of the
single-impurity Anderson model at self-consistency is
derived in~\ref{appA}.

\section{Definitions}
\label{model}

\subsection{Hamiltonian}

We investigate spin-1/2 electrons on a lattice whose 
motion is described by
\begin{equation}
\hat{T} =
\sum_{\sitei,\sitej;\sigma} t_{\sitei,\sitej} 
\hat{c}_{\sitei,\sigma}^+\hat{c}_{\sitej,\sigma} \; ,
\end{equation}
where $\hat{c}^+_{\sitei,\sigma}$,
$\hat{c}_{\sitei,\sigma}$ are creation and annihilation operators for
electrons with spin~$\sigma=\uparrow,\downarrow$ on site~$\sitei$.
The matrix elements
$t_{\sitei,\sitej}$ are the electron transfer amplitudes between
sites $\sitei$~and $\sitej$, and $t_{\sitei,\sitei}=0$.
Since we are interested
in the Mott insulating phase, we consider exclusively a half-filled band
where the number of electrons~$N$ equals the
number of lattice sites~$L$. 

For lattices with translational symmetry, $t_{\sitei,\sitej}=
t(\sitei-\sitej)$,
the operator for the kinetic energy is diagonal in momentum space,
\begin{equation}
\hat{T} = \sum_{\sitek,\sigma} \epsilon(\sitek) 
\hat{c}^+_{\sitek,\sigma}\hat{c}_{\sitek,\sigma}
\; ,
\end{equation}
where
\begin{equation}
\epsilon(\sitek) = \frac{1}{L} \sum_{\sitei,\sitej} t(\sitei-\sitej)
e^{-\I (\sitei-\sitej) \sitek} \; . 
\end{equation}
The density of states for non-interacting electrons is then given by
\begin{equation}
\rho(\epsilon)= \frac{1}{L} \sum_{\sitek} \delta(\epsilon-\epsilon(\sitek))
 \nonumber \;.
\end{equation}
In the limit of infinite lattice dimensions and for translationally
invariant systems without nesting, 
the Hubbard model is characterized by $\rho(\epsilon)$ alone, i.e.,
higher-order correlation functions 
in momentum space factorize~\cite{PvDetal}.
For our explicit calculations we shall later use 
the semi-circular density of states
\begin{eqnarray}
\rho_0(\omega)&=& \frac{2}{\pi W}\sqrt{4 -\left(\frac{4\omega}{W}\right)^2\,}
\quad , \quad   (|\omega|\leq \frac{W}{2}) \; ,
\label{rhozero}\\[3pt]
1 & = & \int_{-W/2}^{W/2} \diff \omega \rho_0(\omega) \; ,
\end{eqnarray}
where $W=4t$ is the bandwidth. In the following, we take $t\equiv 1$
as our unit of energy.
This density of states is realized for non-interacting
tight-binding electrons on a Bethe lattice of connectivity 
$Z\to\infty$~\cite{Economou}.
Specifically, each site is connected to $Z$~neighbors 
without generating closed loops,
and the electron transfer is restricted to nearest-neighbors,
$t_{\sitei,\sitej}=-t/\sqrt{Z}$ when $\sitei$~and $\sitej$~are
nearest neighbors and
zero otherwise. The limit $Z\to\infty$ is implicitly understood henceforth.

The electrons are taken to interact only locally,
and the Hubbard interaction reads
\begin{equation}
\hat{D} = \sum_{\sitei} \left(\hat{n}_{\sitei,\uparrow}-\frac{1}{2}\right)
\left(\hat{n}_{\sitei,\downarrow}-\frac{1}{2}\right) \; ,
\end{equation}
where $\hat{n}_{\sitei,\sigma}=
\hat{c}^+_{\sitei,\sigma}\hat{c}_{\sitei,\sigma}$ 
is the local density operator at site~$\sitei$ for 
spin~$\sigma$. This 
leads us to the Hubbard model~\cite{HubbardI}, 
\begin{equation}
\hat{H}=\hat{T} + U \hat{D} \; .
\label{generalH}
\end{equation}
The Hamiltonian explicitly exhibits particle-hole symmetry,
i.e., $\hat{H}$ is invariant under the particle-hole transformation 
\begin{equation}
\hat{c}_{\sitei,\sigma}^+ \mapsto (-1)^{|\sitei|}
\hat{c}_{\sitei,\sigma}\quad ; \quad
\hat{c}_{\sitei,\sigma} \mapsto (-1)^{|\sitei|} \hat{c}_{\sitei,\sigma}^+ \; ,
\label{phdef}
\end{equation}
where $|{\sitei}|$ counts the number of nearest-neighbor steps
from the origin of the Bethe lattice to site~$\sitei$.
The chemical potential $\mu=0$ then guarantees a half-filled band
for all temperatures~\cite{Buch}.

\subsection{Green Functions}
\label{subsec:Green}

The time-dependent local single-particle Green function at zero temperature
is given by~\cite{Fetter}
\begin{equation}
G(t) = -\I \frac{1}{L} \sum_{\sitei,\sigma} 
\langle \hat{\cal T} [ 
\hat{c}_{\sitei,\sigma}(t)\hat{c}_{\sitei,\sigma}^+] \rangle \; .
\label{GFdef}
\end{equation}
Here $\hat{\cal T}$ is the time-ordering operator and $\langle \ldots \rangle$
implies the average over the degenerate ground states with energy $E_0$,
and (taking $\hbar \equiv 1$ henceforth)
\begin{equation}
\hat{c}_{\sitei,\sigma}(t) = \exp(\I \hat{H} t) \hat{c}_{\sitei,\sigma}
                          \exp(-\I \hat{H} t) 
\end{equation}
is the annihilation operator in the Heisenberg picture. 

In the insulating phase we can readily identify 
the contributions from the lower (LHB)
and upper (UHB) Hubbard bands to the 
Fourier transform of the local Green function ($\eta=0^+$),
\begin{eqnarray}
G(\omega) &=& \int_{-\infty}^{\infty} \diff t e^{\I \omega t} G(t) 
= G_{\rm LHB}(\omega ) + G_{\rm UHB}(\omega ) \; , \nonumber \\
G_{\rm LHB}(\omega ) &=&  \frac{1}{L} \sum_{\sitei,\sigma}
        \left\langle \hat{c}_{\sitei,\sigma}^+
            \left[\omega +(\hat{H}-E_0)-\I\eta\right]^{-1} 
         \hat{c}_{\sitei,\sigma}\right\rangle \; , \nonumber \\ 
G_{\rm UHB}(\omega ) &=& - G_{\rm LHB}(-\omega )  \label{DefGLHB} \; .
\end{eqnarray}
The last equality follows from the particle-hole symmetry~(\ref{phdef}).
Therefore, it is sufficient to
evaluate the local Green function for the lower Hubbard band
which describes the dynamics of a hole inserted into the system.

The density of states for the lower Hubbard band
can be obtained from the imaginary part of
the Green function~(\ref{DefGLHB}) for real arguments via~\cite{Fetter}
\begin{eqnarray}
D_{\rm LHB}(\omega) &=& \frac{1}{\pi} \Im G_{\rm LHB}(\omega)
\; , \nonumber
\\
&=& \frac{1}{L} \sum_{\sitei,\sigma} \left\langle 
\hat{c}_{\sitei,\sigma}^+  \delta\left(\omega+\hat{H}-E_0\right)
\hat{c}_{\sitei,\sigma} \right\rangle \; ,
\label{Dforlateruse}
\label{rangeofD}
\end{eqnarray}
with $ \omega \leq -\Delta(U)/2 <0$, where
$\Delta(U)$ is the single-particle gap.
Particle-hole symmetry results in a symmetric density of states
around $\omega=0$ at half band-filling
\begin{equation}
D(\omega)=D_{\rm LHB}(\omega)+D_{\rm UHB}(\omega)
\label{defDallomega}
\end{equation}
with $D_{\rm UHB}(\omega)=D_{\rm LHB}(-\omega)$. 

We define the (shifted) moments $M_n(U)$ of the density of states 
in the lower Hubbard band via
\begin{equation}
M_n(U)= \int_{-\infty}^{-\Delta(U)/2}
\diff \omega \left(\omega+\frac{U}{2}\right)^n D_{\rm LHB}(\omega) \; .
\label{Mndef}
\end{equation}
In particular, from~(\ref{Dforlateruse}) we find that~\cite{Fetter}
\begin{eqnarray}
M_0(U)&=&1 \; \label{sumrule1}
,\\
M_1(U)&=&
\frac{1}{L}\left(E_0(U)+U\frac{\partial E_0(U)}{\partial U}\right)+\frac{U}{2} 
\label{sumrule2}
\end{eqnarray}
are two useful sum-rules which we shall employ later.
We also note that the average double occupancy is related
to a derivative of the ground-state energy by
\begin{equation}
d(U)= \frac{1}{4}+ \frac{1}{L}\langle \hat{D}\rangle 
=\frac{1}{4} + \frac{1}{L}\frac{\partial E_0(U)}{\partial U} \;.
\label{dbar}
\end{equation}

\subsection{Results from strong-coupling perturbation theory}

We shall test our numerical results against those
from strong-coupling perturbation theory~\cite{mottinsu}.
To second order in $1/U$, the density of states of the lower Hubbard band
reads
\begin{eqnarray}
D_{\rm LHB}(\omega) &=& \int_{-2}^{2} \diff \epsilon \rho_0(\epsilon)
s(\epsilon,U) \delta\left(\omega+\frac{U}{2} +g(\epsilon,U)\right) 
+ {\cal O}(U^{-3}) \; , \nonumber\\
s(\epsilon,U) &=& 1 -\frac{\epsilon}{U} 
+\frac{9\left(\epsilon^2-1\right)}{4 U^2}\; ,
\label{DOSfinalres} \\
g(\epsilon,U) &=& \epsilon 
-\frac{\epsilon^2-3}{2U}
+\frac{3 \epsilon \left(2 \epsilon^2-7\right)}{8 U^2}
\; . \nonumber 
\end{eqnarray}
The zeros of $D_{\rm LHB}(\omega)$ provide
the single-particle gap and the width of the Hubbard bands~$W^*(U)$,
\begin{eqnarray}
\Delta(U)&=& U-4 -\frac{1}{U} -\frac{3}{2 U^2}
+{\cal O}(U^{-3}) \; ,
\label{gapfinal2nd} \\
W^*(U) &=& 4 + \frac{3}{2 U^2} + {\cal O}(U^{-3}) \; ,
\label{width2nd}
\end{eqnarray}
up to second order in $1/U$. The Hubbard bands display a square-root
onset,
\begin{equation}
D_{\rm UHB}(\omega) \sim \left(\omega-\frac{\Delta(U)}{2}
\right)^{1/2} 
\quad , \quad \omega \to \frac{\Delta(U)}{2} \; .
\label{dosexponenta}
\end{equation} 
Note that there is no weight outside the 
Hubbard bands up to and including order $1/U^3$ but there are
contributions to order $1/U^4$ and higher.
Our numerical results show that the weight outside the 
(primary) Hubbard bands at $|\omega|\geq \Delta(U)/2+W^*(U)$
is at most one percent of the
total density of state for all interaction strengths in the insulating
phase.

Recently, E.~Kalinowski~\cite{Bluemer,Eva} 
has calculated the ground-state energy to 11th order in the inverse
coupling strength. Here we restate her results,
\begin{eqnarray}
\fl \hphantom{d(U)} \frac{E_0(U)}{L}  
= - \frac{U}{4} -\frac{1}{2U} -\frac{1}{2U^3}
-\frac{19}{8U^5} -\frac{593}{32U^7}-\frac{23877}{128U^9} 
 -\frac{4496245}{2048 U^{11}}  - {\cal O}(U^{-13}) \; , 
\label{EzeroPT12}\\[3pt]
\fl \hphantom{\frac{E_0(U)}{L}} d(U) 
= \frac{1}{2U^2} +\frac{3}{2U^4}
+\frac{95}{8 U^6} +\frac{4151}{32U^8}+\frac{214893}{128 U^{10}}
+\frac{49458695}{2048 U^{12}}  + {\cal O}(U^{-14}) 
\; . \label{doublePT12}
\end{eqnarray}
Unfortunately, the computational effort 
increases exponentially as a function of the order,
and it will be difficult to obtain much higher orders of the expansion.

\section{Fixed-energy dynamical mean-field theory with
dynamical density-matrix renormalization group [FE-DMFT(DDMRG)]}
\label{method}

In this section, we first discuss the single-impurity model
onto which the Hubbard model can be mapped in the limit
of infinite dimensions. Next, we recall the fixed-energy 
algorithm for the dynamical mean-field theory.
Lastly, we discuss the density-matrix renormalization group
for the numerical solution of the single-impurity Anderson model.

\subsection{Dynamical Mean-Field Theory (DMFT)}
\label{subsec:DMFT}

In the limit of infinite dimensions~\cite{dinfty}, and under the conditions
of translational invariance and convergence of perturbation theory in strong
and weak coupling, the Hubbard model
can be mapped onto single-impurity models~\cite{BrandtMielsch,Jarrell,RMP},
which need to be solved self-consistently. 
In general, these impurity models cannot be solved analytically.

For an approximate numerical treatment various different implementations
are conceivable.
One realization is the single-impurity Anderson model in `star geometry',
\begin{eqnarray}
\hat{H}_{\rm SIAM} &=&\sum_{\ell=1;\sigma}^{n_s-1}
\epsilon_{\ell} \hat{\psi}_{\sigma;\ell}^+\hat{\psi}_{\sigma;\ell}
+ U \left( \hat{d}_{\uparrow}^+\hat{d}_{\uparrow} -\frac{1}{2}\right)
\left( \hat{d}_{\downarrow}^+\hat{d}_{\downarrow} -\frac{1}{2}\right) 
\nonumber \\
&& + \sum_{\sigma} \sum_{\ell=1}^{n_s-1} V_{\ell}
\left( \hat{\psi}_{\sigma;\ell}^+\hat{d}_{\sigma} + 
\hat{d}_{\sigma}^+\hat{\psi}_{\sigma;\ell}\right) \; ,
\label{SIAMns}
\end{eqnarray}
where $V_{\ell}$ are real, positive hybridization matrix elements.
The model describes the hybridization of an impurity site 
with Hubbard interaction to $n_s-1$ bath sites without interaction 
at energies $\epsilon_{\ell}$. 
Here $\hat{d}^+_{\sigma}, \hat{d}_{\sigma}, 
\hat{\psi}^+_{\sigma;\ell}, \hat{\psi}_{\sigma;\ell}$ 
are creation and annihilation operators for
electrons with spin~$\sigma=\uparrow,\downarrow$ on 
the impurity and the bath site~$\ell$, respectively.
In order to ensure particle-hole
symmetry, we have to set $\epsilon_{\ell}=-\epsilon_{n_s-\ell}$
and $V_{\ell}=V_{n_s-\ell}$
for $\ell=(n_s+1)/2,\ldots,n_s-1$. Moreover, since we are interested 
in the Mott--Hubbard insulator, we only use 
odd $n_s$ so that there is no bath state at $\epsilon=0$.

For a given set of parameters $\{\epsilon_{\ell}, V_{\ell}\}$
the model~(\ref{SIAMns}) defines a many-body problem for which the 
single-particle Green function 
\begin{equation}
G_{dd;\sigma}^{(n_s)}(t) = -\I \left\langle \hat{{\cal T}} \left[
\hat{d}_{\sigma}(t) \hat{d}_{\sigma}^+\right]
\right\rangle_{\rm SIAM} 
\label{GSIAMfinite}
\end{equation}
must be calculated numerically.
Here, $\langle \ldots\rangle_{\rm SIAM}$ 
implies the ground-state expectation value within the single-impurity model.

Ultimately, we are interested in the limit $n_s \to\infty$ where 
\begin{equation}
H^{(n_s)}(\omega) 
= \sum_{\ell=1}^{n_s-1} 
\frac{V_{\ell}^2}{\omega-\epsilon_{\ell}+\I 0^+\, \sgn(\omega)} 
\end{equation}
becomes
the hybridization function of the \emph{continuous} problem, 
\begin{equation}
H(\omega) = \lim_{n_s\to\infty}H^{(n_s)}(\omega) 
\end{equation}
and the Green function is  
\begin{equation}
G_{dd}(\omega)
= \lim_{n_s\to\infty}
[G_{dd;\uparrow}^{(n_s)}(\omega) + G_{dd;\downarrow}^{(n_s)}(\omega)]
 \; .
\end{equation}
[For finite $n_s$ the Green functions
$G_{dd;\sigma}^{(n_s)}(\omega)$ are different for 
$\sigma=\uparrow,\downarrow$ because the system contains
an odd number $n_s$ of electrons.]
As shown in~\cite{RMP}, the hybridization function and the Green
function must obey a self-consistency relation,
\begin{equation}
H(\omega) = \frac{G_{dd}(\omega)}{2}
\label{selfcons}
\end{equation}
to describe the Hubbard model on the Bethe lattice 
with connectivity $Z\to\infty$. 
At self-consistency, the Green function of the impurity problem
gives the Green function of the Hubbard model,
\begin{equation}
G_{dd}(\omega) = G(\omega)
\label{selfcons2}
 \; .
\end{equation}
For a finite-size representation of the bath, $n_s< \infty$,
it is generally {\sl not\/} possible to find a self-consistent
solution to the finite-system version of~(\ref{selfcons}),
\begin{equation}
H^{(n_s)}(\omega) = \frac{1}{2}
[G^{(n_s)}_{dd;\uparrow}(\omega) + G^{(n_s)}_{dd;\downarrow}(\omega)]\; .
\label{selfconsfinite}
\end{equation}
Instead, we have to choose bath energies~$\epsilon_{\ell}$
and hybridizations~$V_{\ell}$ for finite~$n_s$
in such a way that the
single-particle Green function and 
the hybridization function fulfill~(\ref{selfcons}) for
$n_s\to\infty$. Therefore, numerical methods 
will differ in the way an approximate self-consistency 
condition is defined.
This is a source of ambiguity because
there can be more than one self-consistent set of parameters 
$\{\epsilon_{\ell}, V_{\ell}\}$ for fixed $n_s$.
Moreover, it cannot be guaranteed that different schemes
will ultimately coincide for $n_s \to\infty$.

\subsection{Fixed-energy dynamical mean-field theory (FE-DMFT)}
\label{subsec:FEimplementation}

In Ref.~\cite{mottinsu} a new algorithm
for solving the self-consistency problem has been introduced.
The accuracy and stability of this `fixed-energy DMFT' approach
has been demonstrated using an exact diagonalization technique
as `impurity solver', i.e., to compute the single-impurity
Green function~$G_{dd,\sigma}^{(n_s)}(\omega)$.
In this work, we describe how to use the 
FE-DMFT
together with the dynamical density-matrix renormalization group
as impurity solver.

For the Mott--Hubbard insulator, we make the assumption
that all the spectral weight is concentrated in the upper and lower
Hubbard bands, i.e., in the finite frequency interval 
\begin{equation}
I=\left\{\omega \ \Bigl| \ 
\frac{\Delta(U)}{2} \leq |\omega| \leq \frac{\Delta(U)}{2} +W^*(U)
\right\}
\label{freqninterval}
\; .
\end{equation}
The onset of the upper Hubbard band, $\Delta(U)/2$,
and the width of the Hubbard bands, $W^*(U)$, are determined
self-consistently; see below.  
We start with some input guess $\Delta(U)$ 
and $W^*(U)$, 
which we may take from second-order perturbation theory~(\ref{gapfinal2nd}),
from the fixed-energy dynamical mean-field theory 
with exact diagonalization~\cite{mottinsu},
or from previous runs for slightly different values of~$U$ or~$n_s$.
We discretize the Hubbard bands equidistantly,
i.e., we fix the energies 
$\epsilon_{\ell}$ by
\begin{equation}
\epsilon_{\ell}= -\epsilon_{n_s - \ell} = \frac{\Delta(U)}{2} 
+ \left(\ell-\frac{1}{2}\right) \delta(U) 
\quad , \quad 1\leq \ell \leq (n_s-1)/2 \; ,
\label{discretization}
\end{equation}
where 
\begin{equation}
\delta(U)= \frac{2W^*(U)}{n_s-1}
\end{equation}
is the distance between
two consecutive energies $\epsilon_{\ell}$ in the same Hubbard band. 
Then we divide the interval $I$ into $n_s-1$ intervals
$I_{\ell}$ of width $\delta(U)$ centered 
around each energy $\epsilon_{\ell}$.
By fixing the energies at the centers of equidistant intervals 
we can be sure that our
resolution of the Hubbard bands becomes increasingly better as $n_s$ increases.
For a typical $n_s=65$ and $W^*(U) \approx 4t$
we have $\delta(U)\approx 0.125$.

When we integrate the imaginary part of the Green function
over the interval $I_{\ell}$
we obtain weights $w_{\ell}$,
\begin{equation}
w_{\ell}
= \int_{I_{\ell}} \diff \omega \frac{|\Im 
 G_{dd}(\omega)|}{2} 
\; .
\label{recollectionofweight}
\end{equation}
At self-consistency~(\ref{selfcons}) and for $n_s\to\infty$, 
these weights obey
\begin{equation}
V_{\ell}^2=w_{\ell} \; .
\label{closefirstloop}
\end{equation}
We can use this relation to calculate new parameters~$V_{\ell}$ from  
a Green function~$G_{dd}(\omega)$.
As initial values for $G_{dd}(\omega)$
we may again use the result of second-order 
perturbation theory~(\ref{DOSfinalres}) in~(\ref{recollectionofweight}),
the results of the 
FE-DMFT(ED)~\cite{mottinsu},
or we start from previous runs for slightly different values of~$U$ or~$n_s$.
The latter approach is recommendable close to the transition.

At every iteration the impurity Green function
$G_{dd}(\omega)$ must be calculated
with the help of an `impurity solver'.
Here, we use the dynamical DMRG to calculate 
$G_{dd,\sigma}^{(n_s)}(\omega)$ for the Hamiltonian~(\ref{SIAMns})
with finite $n_s$.
Then, the deconvolution of the sum of these Green functions for 
$\sigma = \uparrow, \downarrow$
gives an excellent approximation of the 
Green function $G_{dd}(\omega)$ in the limit $n_s \rightarrow \infty$
at all needed frequencies (see the next subsection).

We now describe the iterative procedure used
to determine the onset of the upper Hubbard band $\Delta(U)/2$,
its width~$W^*(U)$, and the Green functions $G_{dd}(\omega)$
self-consistently.  
Starting from the initial $\Delta(U)/2$, $W^*(U)$, and
$G_{dd}(\omega)$, we compute
the energies $\epsilon_{\ell}$ and hybridization
matrix elements $V_{\ell}$ of the single-impurity Anderson 
model~(\ref{SIAMns}) using equations~(\ref{discretization})
to (\ref{closefirstloop}).
In a first calculation we consider this model with $n_s$ sites
and use the DDMRG method to compute the full Green functions 
$G_{dd,\sigma}^{(n_s)}(\omega)$ 
with a resolution $\eta \sim \delta(U) \sim 1/n_s$.
As explained above, after deconvolution of these Green functions we 
obtain a new Green 
function $G_{dd}(\omega)$, which is used in the next iteration.
Simultaneously, we use the DDMRG method with a broadening
$\eta \ll \delta(U)$
to compute the energy $\Delta(U,n_s')/2$
of the first pole in $G_{dd,\sigma}^{(n_s)}(\omega)$
(i.e., the lowest state contributing to the density of states)
for the single-impurity Anderson model with $n_s' \geq n_s$ sites.
Typically, we calculate~$\Delta(U,n_s')$
for $n_s'=81,97,113,129,145,161$.
[These calculations can be carried out for larger system sizes
than the calculation of the full Green functions because
we only need to determine ground-state properties and a small fraction 
of the Green function spectrum around $\omega \approx \Delta(U)/2$.]
After extrapolating to the thermodynamic limit, 
\begin{equation}
\Delta(U) = \lim_{n_s' \to\infty}\Delta(U,n_s')
\; , 
\label{deltagsns}
\label{energycriterion}
\end{equation}
we obtain a new estimate for the onset of the upper Hubbard band,
$\Delta(U)/2$, which is used in the next iteration. 
At the same time we use the sum rule of~\ref{appA} 
for the ground-state energy $E_0^{\rm SIAM}(U,n_s')$ of the effective
single-impurity Anderson model to calculate a new bandwidth,
\begin{equation}
\frac{\Delta(U)+W^*(U)}{2} = \lim_{n_s'\to\infty} 
\frac{E_0^{\rm SIAM}(U,n_s')}{n_s'} \; .
\label{eqfromappA}
\end{equation}

After a new gap~$\Delta(U)$, bandwidth~$W^*(U)$, and
Green function $G_{dd}(\omega)$ have been obtained
we can start the next iteration.
We repeat this procedure until it converges to a fixed point.
Typically, we need less than 10~iterations for the procedure
to converge, depending on the choice of the starting parameters.
We terminate the iterative procedure when the variation of 
the gap~$\Delta(U)$ and bandwidth~$W^*(U)$ is smaller than 
$10^{-3}t$ from one iteration to the next.
At that point the variation of $G_{dd}(\omega)$ is found to be smaller than
$10^{-3}$ for all frequencies $\omega$.
This iterative procedure is stable;
for small deviations from the self-consistent values,
the gap and the width of the Hubbard bands are driven back
to the fixed point of the iteration.
We have also checked that, for fixed $n_s$, 
a unique solution for $G_{dd}(\omega)$ is 
found for various starting choices. 
Obviously our FE-DMFT(DDMRG) 
approach yields self-consistent results
for the gap, bandwidth, and Green function of the Hubbard
model. Moreover, it is possible to calculate ground-state properties
of the Hubbard model (energy, double occupancy)
from ground-state properties of the
self-consistent single-impurity Anderson model, as shown in the 
next section.

\begin{figure}[htb]
\begin{center}
\resizebox{10cm}{!}{\includegraphics{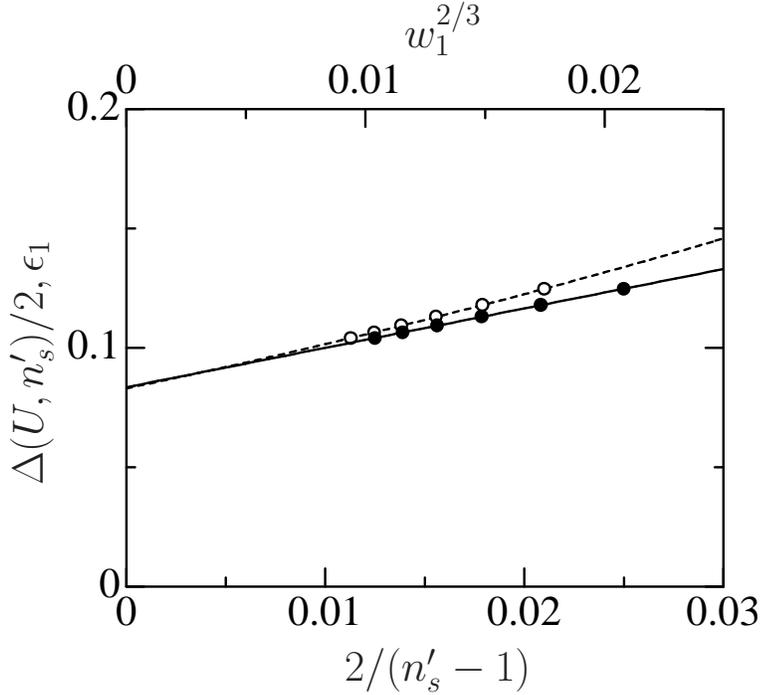}}
\caption{Extrapolation of the lowest lying single-particle excitation energy
$\Delta(U,n_s')/2$
as a function of inverse system size (lower axis)
for $U=4.6$ at self-consistency (solid circles). 
The open circles give the energy $\epsilon_1$ as a function of the
weight $w_1$ (upper axis) for the same system sizes
$n_s'=81,97,113,129,145,161$.
\label{Fig:convergence}}
\end{center}
\end{figure}

In Figure~\ref{Fig:convergence} we give an example
of the extrapolation scheme for $\Delta(U,n_s')$ at
the fixed-point of our iterative procedure for $U=4.6$.
As expected, the results for $81 \leq n_s' \leq 161$ extrapolate 
linearly in $1/n_s'$.
Note that the FE-DMFT with exact diagonalization~\cite{mottinsu} 
was limited to $n_s'=n_s=15$, and finite-size effects 
had to be controlled
by a combination with the criterion of a square-root onset
of the Hubbard bands, which is 
suggested by perturbation theory~(\ref{dosexponenta}).
The DDMRG treats 
system sizes up to $n_s' \sim {\cal O}(200)$ which makes
this `weight criterion' obsolete. 
Nevertheless, we can use the `weight criterion' as 
a consistency check. As argued in~Ref.~\cite{mottinsu},
we should find
\begin{equation}
\epsilon_1 - \frac{\Delta(U,n_s')}{2} 
\propto \left(w_1\right)^{2/3}  \; ,
\label{weightcriterion}
\end{equation}
for a square-root increase of the density of states near the band
edges. 
In Fig.~\ref{Fig:convergence} we show
$\epsilon_1$ as a function of $w_1^{2/3}$ for system sizes 
$81 \leq n_s' \leq 161$
as open circles. Both extrapolated values for the gap
from~(\ref{energycriterion}) and~(\ref{weightcriterion}) agree.
The linear behavior of $\epsilon_1$ as a function of
$w_1^{2/3}$ confirms the square-root increase
of the density of states near the gap.
Note, however, that the region in which the square-root onset
is discernible becomes very small close to the transition and 
thus large system sizes are required.

For $U\leq 4.6$, a constant discretization of the Hubbard band
with $\delta(U)\approx0.125$  
is not sufficient to resolve fine structures of the density of states
near the single-particle gap.
In order to obtain a better
resolution for  $|\omega| \approx \Delta(U)/2$ without
excessive increase of the computational effort
we use a variable discretization scheme as described in
Ref.~\cite{EricSatoshi}.
The resolution around the gap is improved by 
using a finer discretization $\delta(U)$ of the intervals
$\Delta(U)/2 < |\omega| < t$ (i.e., more  
bath states are used in those intervals).
The smaller $\delta(U)$ allows us to use
a smaller broadening $\eta$ in DDMRG calculations for
those frequencies.
We combine the high-energy spectrum obtained with 
the usual resolution
and the low-energy spectrum obtained
with the improved resolution 
and then deconvolve the result to obtain a new 
Green function $G_{dd}(\omega)$.
This yields $G_{dd}(\omega)$ for $|\omega| < 0.6t$ with a resolution, 
which is up to an order of magnitude better than for a constant
discretization with $n_s =65$. 
For the results presented here 
we have used  $\delta(U) = 0.02$ (corresponding to
$n_s=113$ and $\eta =0.03$)  
for $U=4.5$ and $\delta(U) = 0.031$ ($n_s=97$ and $\eta = 0.05$)
for $U=4.6$ in the intervals $\Delta(U)/2 < |\omega| < t$.

\subsection{DDMRG for the single-impurity Anderson model}

The DDMRG for the single-impurity Anderson model
is described in detail in Ref.~\cite{EricSatoshi}.
Here, we summarize the essential ingredients.
As the DMRG method is most accurate for systems with a quasi one-dimensional 
structure, we perform calculations 
of the single-impurity Anderson model~(\ref{SIAMns})
in its equivalent linear-chain form~\cite{Costi}
\begin{eqnarray}
\fl
\hat{H}_{\rm SIAM} &=&
 U \left( \hat{d}_{\uparrow}^+\hat{d}_{\uparrow} -\frac{1}{2}\right)
\left( \hat{d}_{\downarrow}^+\hat{d}_{\downarrow} -\frac{1}{2}\right) 
+ V \sum_{\sigma} \left( \hat{f}_{\sigma;0}^+\hat{d}_{\sigma} +
\hat{d}_{\sigma}^+\hat{f}_{\sigma;0}\right) 
\label{SIAMchain}  \\
\fl
&& + \sum_{\sigma} \sum_{\ell=0}^{n_s-2} \lambda_{\ell}
\left( \hat{f}_{\sigma;\ell}^+\hat{f}_{\sigma;\ell+1} + 
\hat{f}_{\sigma;\ell+1}^+\hat{f}_{\sigma;\ell}\right) \; . \nonumber
\end{eqnarray}
The DDMRG provides the local density of states 
\begin{equation}
D_{dd,\sigma}^{\eta}(\omega_i)
= -\sgn (\omega_i) \frac{\Im G_{dd,\sigma}^{(n_s)}(\omega_i)}{\pi}
\label{DOSdef}
\end{equation}
at selected frequencies~$\omega_i$ very accurately. 
The real part of the Green function can also be calculated 
with DDMRG but
to carry out the FE-DMFT calculation we need only the imaginary
part.
To simulate the continuous spectrum of an 
infinite chain in a calculation with a finite~$n_s$, 
a broadening~$\eta$ is introduced which must be scaled as a function
of the system size~\cite{Eric}. 
If $\eta$ is chosen too small, the density of states displays 
finite-size peaks as those seen in Ref.~\cite{RozenbergHallberg}. 
If $\eta$ is chosen too large, relevant
information is smeared out. As an empirical fact, 
$\eta \sim \delta(U)=2W^*(U)/(n_s-1)$ should be chosen, 
i.e., the resolution
scales as the inverse system size, as found for one-dimensional 
lattice models~\cite{Eric}. 

\begin{figure}[htb]
\begin{center}
\resizebox{12cm}{!}{\includegraphics{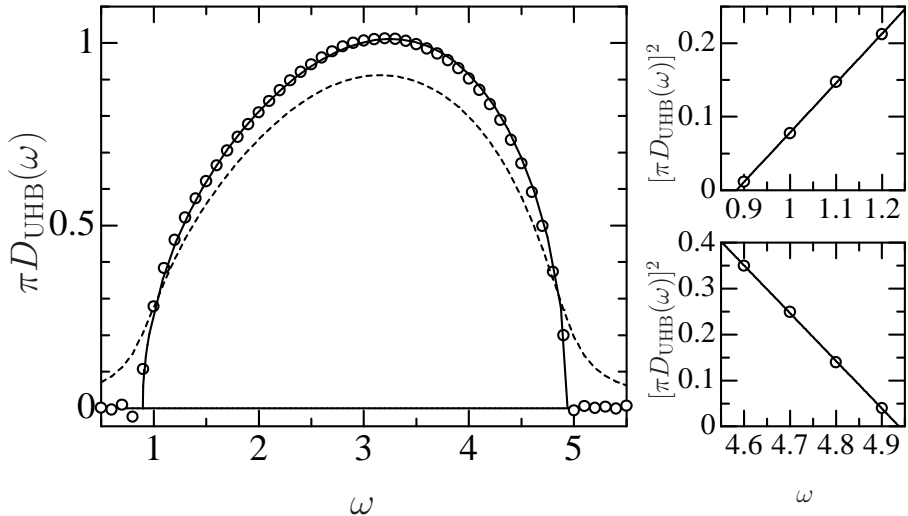}}
\caption{Main figure: density of states of the upper Hubbard band for $U=6$.
The dashed line shows the result of the DDMRG with a broadening
$\eta=0.2$. The circles give the DDMRG density of states
after deconvolution ($\eta=0$).
The full line is the result from second-order perturbation theory
in $1/U$~(\protect\ref{DOSfinalres}). 
The side figures show the linear behavior of the square of the density of 
states as a function of frequency near the band edges.
The lines are linear fits.
\label{Fig:DOSU6}}
\end{center}
\end{figure}

In order to carry out 
the iterative procedure  described in the previous section, 
we determine the density of states at selected
frequencies $\omega_i$. Typically, we choose them to resolve the effective
bandwidth~$W^*(U)$ equidistantly, $\omega_{i+1}-\omega_i= \delta \omega 
\approx \eta \sim \delta(U) $.
We then `deconvolve' the DDMRG data by inverting the 
Lorentz transformation~\cite{mottmetal}
\begin{equation}
D_{dd}^{\eta}(\omega_i) = \sum_{j} \frac{\delta \omega}{\pi}
\frac{\eta}{\eta^2 + (\omega_i-\omega_j)^2} 
D_{dd}(\omega_j) 
\; ,
\end{equation}
where $D_{dd}^{\eta}(\omega) = D_{dd,\uparrow}^{\eta}(\omega) + 
D_{dd,\downarrow}^{\eta}(\omega)$.
Through equation~(\ref{DOSdef})
this deconvolved density of states $D_{dd}(\omega)$
determines the imaginary part of the Green function $G_{dd}(\omega)$ 
which is used in the FE-DMFT(DDMRG) 
scheme.
The procedure can be repeated for different choices of the equidistant
frequencies~$\omega_j$ to get more values of $D_{dd,\sigma}(\omega_j)$.
In practice, we use two
different sets of frequencies, corresponding to a frequency resolution
comparable to the bath energy resolution $\delta(U)$. 
In this way, DDMRG provides a set of values $D_{dd,\sigma}(\omega_j)$
for the density of states. 
The main advantage of this deconvolution 
is that no extrapolation or finite-size scaling
analysis of these values $D_{dd,\sigma}(\omega_j)$ is necessary because they
converge very quickly to the $n_s \to \infty$ limit.
Naturally, structures with an intrinsic width 
of less than $\eta$ cannot be resolved with this procedure even if
we use many different sets of frequencies. 
Therefore, with DDMRG we obtain an accurate discrete representation
of the density of states for the continuum model
[and thus of the imaginary part of $G_{dd}(\omega)$], 
except for small regions 
around its onset and closing points where the derivative of
the density of states changes singularly.
With the DDMRG method~\cite{Eric}, we calculate the one-particle Green 
function~(\ref{GSIAMfinite}) for system sizes up to 
$n_s \sim {\cal O}(200)$.
Therefore the FE-DMFT(DDMRG) leads to a much better resolution
of the Hubbard bands than our previous FE-DMFT with
exact diagonalization which was limited to $n_s=15$.

An example of the density of states obtained with the
FE-DMFT(DDMRG) approach is shown in Figure~\ref{Fig:DOSU6} for $U=6$.
For this interaction strength, the agreement of the deconvolved
DDMRG data with the second-order strong-coupling
perturbation theory~(\ref{DOSfinalres}) for
the Hubbard model is almost perfect.
Our deconvolution scheme gives slightly negative values 
in the vicinity of the band edges. These effects are small and
are to be expected for sharp band edges 
in the density of states at $\omega=\Delta(U)/2$
and $\omega=\Delta(U)/2+W^*(U)$. 
We note that our numerical results are in much better agreement
with perturbation theory than the results
obtained in a recent DMFT(DMRG) study~\cite{RozenbergHallberg} 
where Lanczos-DMRG and a different
self-consistency scheme has been used.
Therefore, we think that our results 
for the gap and the critical interaction strength
are also more accurate than those presented in 
that work~\cite{RozenbergHallberg}.

\section{Results}
\label{results}

In this section we present the results for the Mott
insulating phase of the Hubbard model which we have
obtained with our FE-DMFT(DDMRG) approach.
For ground-state properties
comparisons with strong-coupling perturbation 
theory~\cite{mottinsu,Bluemer,Eva}
and DMFT(QMC) results
(extrapolated to zero temperature)~\cite{Bluemer,Bluemer2}
confirm the accuracy of our method.
Moreover, we will present results for the 
(zero-temperature) single-particle excitations which
are much more accurate than those obtained with 
other DMFT approaches. 

\subsection{Ground-state properties}

\begin{figure}[htb]
\begin{center}
\resizebox{9cm}{!}{\includegraphics{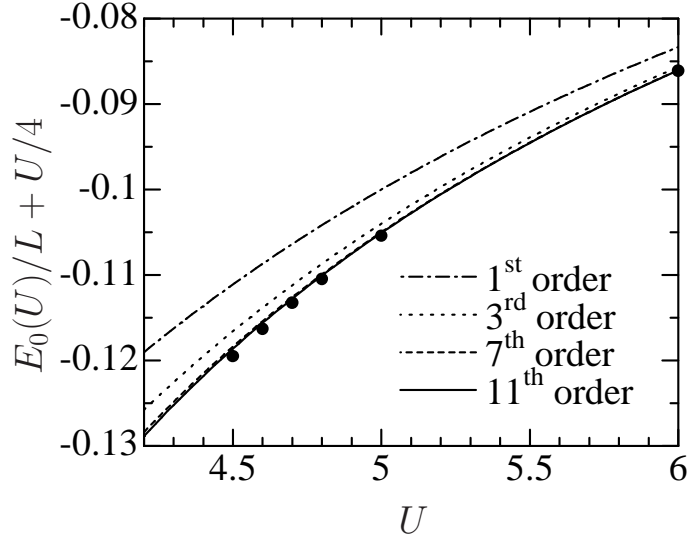}}
\caption{Ground-state energy $E_0/L+U/4$
of the Mott--Hubbard insulator as a function
of the interaction strength.
FE-DMFT(DDMRG) results for $U=4.5,4.6,4.7,4.8,5,6$ (circle) and
perturbation theory (lines) for various orders in $1/U$.
\label{Fig:energy}}
\end{center}
\end{figure}

The ground-state energy per site of the Hubbard model
can be calculated from ground-state expectation values of 
the self-consistent single-impurity Anderson model (see~\ref{appA})
\begin{equation}
\frac{E_0(U)}{L} + \frac{U}{4} = U \langle 
\hat{d}_{\uparrow}^+\hat{d}_{\uparrow} 
\hat{d}_{\downarrow}^+\hat{d}_{\downarrow} \rangle_{\rm SIAM}
+ V \langle \hat{d}_{\sigma}^+\hat{f}_{0,\sigma} \rangle_{\rm SIAM} \; ,
\label{energy}
\end{equation}
where the two terms on the right-hand-side are the interaction 
and kinetic energy per site, respectively, and $V=t=1$.
In Figure~\ref{Fig:energy} we show the ground-state energy
$E_0(U)/L+U/4$ in the Mott--Hubbard insulator phase  
for $4.5 \leq U \leq 6$
in comparison with strong-coupling perturbation theory~(\ref{EzeroPT12}). 
We see that there is a very good agreement between our numerical
data and the analytical results.
Our data points 
lie below the best perturbative energy
(11th-order in $1/U$).
As expected, deviations from the perturbative results
become larger when $U$ becomes smaller, from about $0.8 \times 10^{-4}$ at
$U=6$ to $8.8 \times 10^{-4}$ at $U=4.5$.
Our FE-DMFT(DDMRG) energies are also lower than
DMFT(QMC) energies~\cite{Bluemer,Bluemer2}. However, the differences 
are small, of the order of $2 \times 10^{-4}$ or less, for
$U \geq 4.8$.  
As the Mott insulator solution disappears for
$U < 4.8$ in the DMFT(QMC) approach, no comparison 
with our data is possible below that coupling strength.

\begin{figure}[htb]
\begin{center}
\resizebox{12cm}{!}{\includegraphics{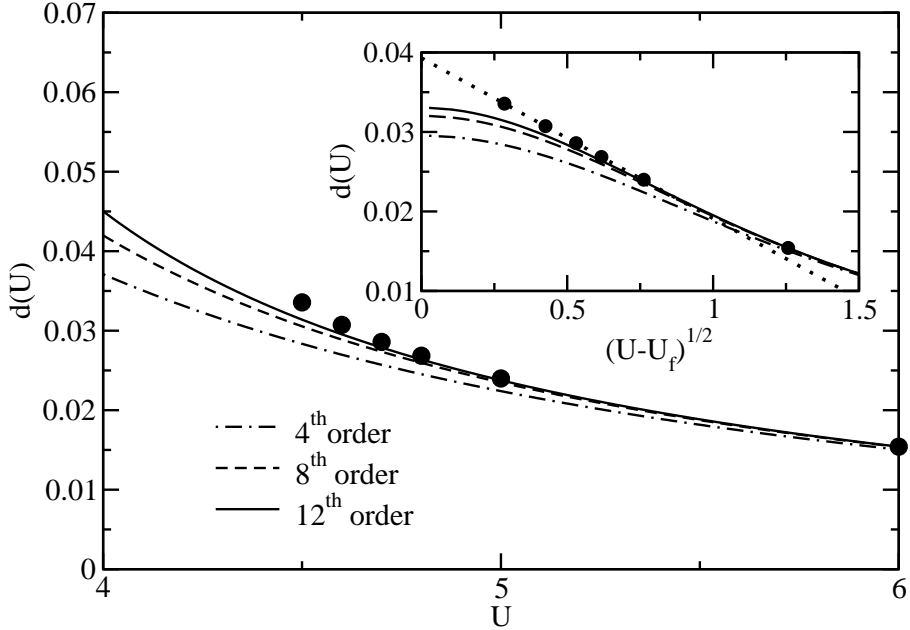}}
\caption{Average double occupancy in the Mott--Hubbard insulator as 
a function of the interaction strength $U$. 
FE-DMFT(DDMRG) results for $U=4.5,4.6,4.7,4.8,5,6$ (circle) and 
perturbation theory for various orders in $1/U$ (lines).
Inset: same results as a function of $(U-U_{\rm f})^{1/2}$ with
$U_{\rm f}=4.419$. The dotted line is a fit to~(\ref{double_scaling}).
\label{Fig:double}}
\end{center}
\end{figure}

It is difficult to evaluate the relative accuracy of our method
from the ground-state energy alone because that quantity is
only defined up to a constant.
The average double occupancy of the Hubbard model is given
by the average double occupancy of the impurity site
in the single impurity Anderson model at self-consistency
\begin{equation}
d(U) = \langle 
\hat{d}_{\uparrow}^+\hat{d}_{\uparrow} 
\hat{d}_{\downarrow}^+\hat{d}_{\downarrow} \rangle_{\rm SIAM} \;.
\label{doubly}
\end{equation}
At half filling this quantity takes only values between zero and 1/4
and thus provides a good benchmark.
In Figure~\ref{Fig:double} we compare our 
FE-DMFT(DDMRG) results
for the average double occupancy with  
perturbation theory~(\ref{doublePT12}) up to 12th~order in $1/U$. 
Again, the agreement is very good but deviations become
clearly noticeable for $U < 5$.
Quantitatively, the differences between our values for $d(U)$ 
and the results of the 12th-order perturbation expansion increase 
significantly from 
$2 \times 10^{-6}$ (about 0.01\%) at $U=6$ to 
$2 \times 10^{-3}$ (about 7\%) at $U=4.5$. 
This is not surprising because we locate the critical value 
below which the Mott insulator no longer exists at 
$U_{\rm c} \approx 4.45 \pm 0.05$ (see below). 
The series expansion for the ground-state energy~(\ref{EzeroPT12})
and the 
average double occupancy~(\ref{doublePT12})
converges only for $U >U_{\rm c}$.
Therefore, the results of finite-order perturbation theory
rapidly become inaccurate as $U$ approaches $U_c$.
A comparison between FE-DMFT(DDMRG) and 
DMFT(QMC)~\cite{Bluemer,Bluemer2}
data is more conclusive.
Both approaches provide
results for the average double occupancy 
which deviate from each other by less than $3 \times 10^{-5}$, corresponding
to relative errors of the order of  0.1\%, down to $U=4.8$.

With our FE-DMFT(DDMRG) approach the Mott insulator is stable for 
significantly weaker couplings than
predicted by other works~\cite{NRG,Bluemer}.
A closer inspection of our data for small $U \leq 5$ shows that
the double occupancy scales as 
\begin{equation}
d(U) = d_{\rm f} - C \sqrt{U-U_{\rm f}} \;.
\label{double_scaling}
\end{equation}
This behavior is clearly seen in the inset of Fig.~\ref{Fig:double}.
A fit of our  data for $U < 5$ gives $U_{\rm f} =4.419$,
$d_{\rm f}=0.03931$ and $C= 0.0202$.
Equation~(\ref{double_scaling}) suggests that the double occupancy is
a singular function of the coupling $U$ at $U_{\rm f}$. 
It is thus reasonable to identify $U_{\rm f}$ with the critical coupling
below which the Mott insulator no longer exists.
The value $U_{\rm f} =4.419$ is indeed in very good agreement with 
the coupling $U_{\rm c} =4.45 \pm 0.05$ where the Mott gap $\Delta(U)$
closes (see below). 
As the average double occupancy is related to the ground-state
energy by~(\ref{dbar}), one expects that
\begin{equation}
\frac{E_0(U)}{L}+\frac{U}{4} = e_0 + d_{\rm f} (U-U_{\rm f}) 
- \frac{2C}{3} (U-U_{\rm f})^{\frac{3}{2}} 
\label{energy_scaling}
\end{equation}
for $U \rightarrow U_{\rm f}$.
Our data for the ground-state energy for $U < 5$
are reproduced by this formula within $5 \times 10^{-5}$
if we use $e_0 = 0.12235$ and the parameters $d_{\rm f},U_{\rm f}$ and $C$
determined from the fit of the double occupancy data.
Therefore, our FE-DMFT(DDMRG) data for the ground-state
energy and the average double occupancy of the Hubbard
model with $4.5 \leq U < 5$
fulfill the relation~(\ref{dbar}) very precisely.
For an arbitrary single-impurity Anderson model
the derivative of the expectation value in the right-hand-side
of~(\ref{energy}) is not given by the average double 
occupancy~(\ref{doubly}).
The relation~(\ref{dbar}) between ground-state energy and double
occupancy is valid for the Hubbard model and thus only for the
expectation values~(\ref{energy}) and~(\ref{doubly}) of
single-impurity Anderson model \textit{at self-consistency}.
Therefore, the scalings~(\ref{double_scaling}) and~(\ref{energy_scaling})
of our data confirms that we have found self-consistent DMFT
solutions~(\ref{selfcons}) for the Hubbard model
with couplings $4.5 \leq U < 5$.

Moreover, these results show that 
our FE-DMFT(DDMRG) approach is accurate enough
to allow for an analysis of the critical behavior
and to determine critical exponents
for the ground-state energy and the average double occupancy.
Note that the behavior~(\ref{double_scaling})
of the average double occupancy implies
that the interaction energy $Ud(U)$ scales as
$U_{\rm f} d_{\rm f} -C U_{\rm f} \sqrt{U-U_{\rm f}}$ 
close to $U_{\rm f}$ and, consequently,
the kinetic energy scales as 
$K +C U_{\rm f} \sqrt{U-U_{\rm f}}$ for small $U-U_{\rm f}$,
where $e_0 = K + U_{\rm f} d_{\rm f}$. 
Recently, evidence for half-integer critical exponents
have also been found using an analysis of the strong-coupling 
perturbation theory extrapolated to infinite order~\cite{Bluemer}.
However, the first singular terms in the expansions of $E_0(U)$ and
$d(U)$  were found to have exponents $5/2$ and $3/2$, respectively,
compared to $3/2$ and $1/2$ in~(\ref{energy_scaling})
and~(\ref{double_scaling}).

\begin{figure}[htb]
\begin{center}
\resizebox{9cm}{!}{\includegraphics{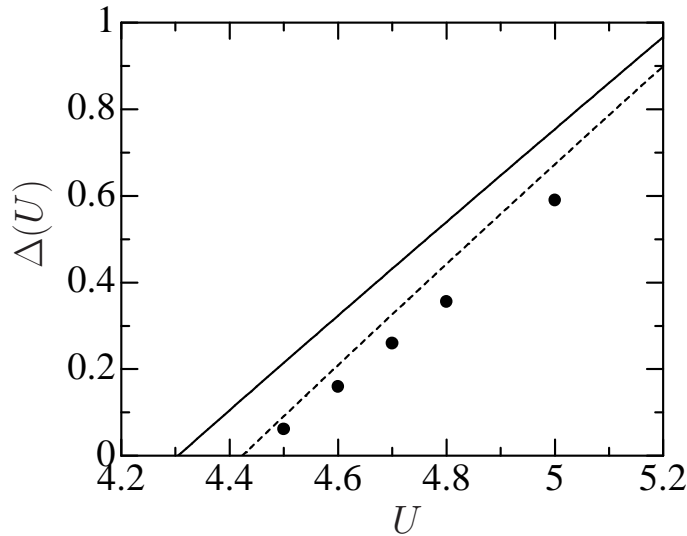}}
\caption{Single-particle gap in the Mott--Hubbard insulator as a function
of the interaction strength.
FE-DMFT(DDMRG) results (circle), 
second-order perturbation theory (solid line),
and the interpolated result from FE-DMFT(ED) (dashed line). 
\label{Fig:gap}}
\end{center}
\end{figure}

\subsection{Single-particle excitations}

The single-particle gap $\Delta(U)$ found at the fixed-point
of our iterative procedure is shown in Figure~\ref{Fig:gap}.
As expected, $\Delta(U)$ first decreases monotonically with $U$
then vanishes below a finite $U_{\rm c} >0$.
For $U=4.5$ the gap $\Delta(U) = 0.062$ is still large enough to be 
detected with our method but for $U=4.4$ we find $\Delta(U) = 0$. 
Thus we estimate that $U_{\rm c} \approx 4.45$ with an error smaller
than $0.05$ in full agreement with the singular behavior of the
ground-state energy~(\ref{energy_scaling})
and double occupancy~(\ref{double_scaling}) described above.
In Fig.~\ref{Fig:gap} it is seen that second-order perturbation
theory describes the gap behavior qualitatively
but it predicts a vanishing of the gap at a slightly too small
$U_{\rm c}=4.31$. 
We also see that our FE-DMFT(DDMRG) results agree with the
results from the FE-DMFT(ED) investigation~\cite{mottinsu}.
The small deviations are within the error estimates
for FE-DMFT(ED) calculations (see Ref.~\cite{mottinsu}).
Concomitantly, the values for
the closing of the gap are almost equal, $U_{\rm c}= 4.43 \pm 0.05$
with the FE-DMFT(ED) method.

Our result for $U_{\rm c}$ is in conflict with the
value $U_{\rm c}=4.78$
found using a DMFT(NRG) approach~\cite{NRG}
and an analysis of the strong-coupling expansion~\cite{Bluemer}.
In contrast, we find substantial gaps $\Delta(U=4.8)= 0.356$ 
and $\Delta(U=4.7)=0.260$ just above and below that coupling.
These gaps are clearly larger than 
the discretization of the bath $\delta(U)=0.125$ that we have used. 
Therefore, we are confident that $U_{\rm c}<4.7$, 
and that our result
$U_{\rm c}\approx 4.45$ is more reliable than the results of
Refs.~\cite{NRG,Bluemer}. 

For large interaction strengths, the derivative of the gap~$\Delta(U)$ 
with respect to~$U$ is unity, $\Delta'(U\gg W)=1$,
see~(\ref{gapfinal2nd}). For finite~$U$, our results show that
$\Delta'(U)>1$, in agreement with perturbation
theory~(\ref{gapfinal2nd}). In the vicinity of the transition,
$U\approx U_{\rm c}$, $\Delta'(U)$ again approaches unity and
thus $\Delta(U) = U-U_{\rm c}$.

The width of the Hubbard bands 
$W^*(U)$ calculated at the fixed-point of the
FE-DMFT(DDMRG) procedure is almost
constant for all $U > U_{\rm c}$. At finite coupling it is
slightly larger than the value $W^*(U) = W =4$ 
predicted by strong-coupling perturbation theory 
for $U\rightarrow \infty$ and
reaches a maximum $W^*(U) \approx 4.1$
around $U=5$.

\begin{figure}[htb]
\begin{center}
\resizebox{12cm}{!}{\includegraphics{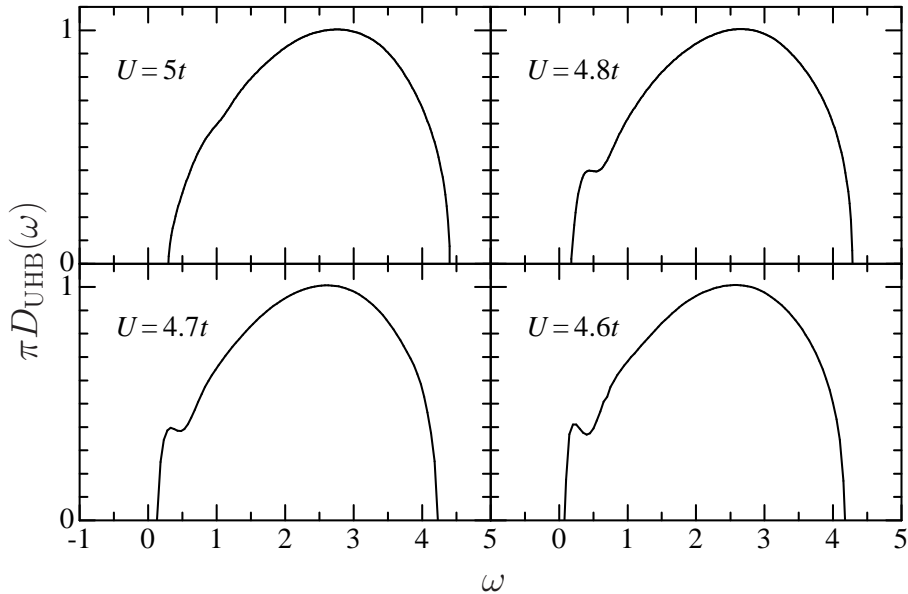}}
\caption{Density of states of the upper Hubbard band for 
$4.6 \leq U \leq 5$.
\label{Fig:DOSUx}}
\end{center}
\end{figure}

In order to explain this behavior, we display the density
of states as a function of~$U$ in Fig.~\ref{Fig:DOSUx}.
For large interaction strengths, $U\geq 6$,
second-order perturbation theory describes
the density of states $D(\omega)$
accurately, as seen in Fig.~\ref{Fig:DOSU6}.
The spectrum consists of the two
Hubbard bands around $\pm U/2$ with
a square-root onset at $\omega = \pm \Delta(U)/2$.
For weaker couplings our FE-DMFT(DDMRG) calculations
show clearly that a shoulder forms in the density of states
near the transition to the metallic phase.
In Fig.~\ref{Fig:DOSUx}, we can see that this
feature becomes progressively stronger as $U$~approaches
$U_{\rm c}$. Its appearance is connected with
the non-monotonous behavior of~$W^*(U)$ and of~$\Delta'(U)$
as a function of~$U$ near $U=5$. 

\begin{figure}[htb]
\begin{center}
\resizebox{10cm}{!}{\includegraphics{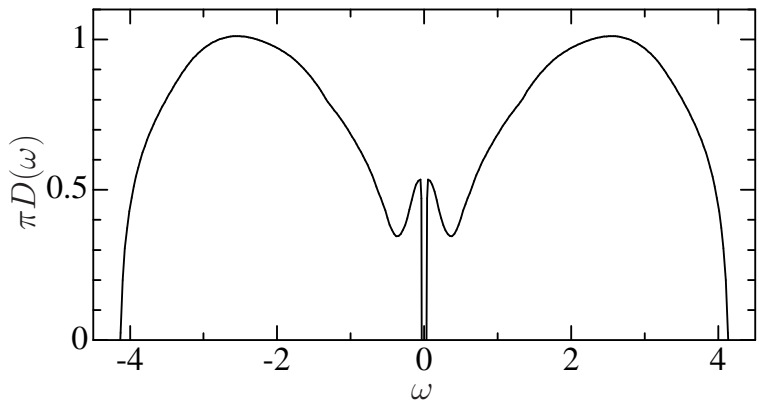}}
\caption{Density of states of the lower and upper Hubbard bands for $U=4.5$.
\label{Fig:DOSU4c5}}
\end{center}
\end{figure}

This feature is the remainder
of the quasi-particle peak which is present at $\omega=0$
in the metallic phase. 
Because the metal is a Fermi liquid,
the quasi-particle peak 
has height $D(\omega=0)=\rho(0)=1/\pi$~\cite{MHdinfty}.
Figure~\ref{Fig:DOSU4c5} suggests that the quasi-particle peak splits
at the transition to the insulating state at $U_{\rm c}$.
(A splitting of the density of states at the 
transition also occurs in the one-dimensional Hubbard model
where $U_{\rm c}=0^+$, as shown in Ref.~\cite{Fabianoned}
within a field-theoretical approach.)
As the gap opens, the two flanks of the peak quickly loose weight
so that they are rather small already at $U=4.5$.

Clearly, our FE-DMFT(DDMRG) results 
for the gap and the bandwidth are more accurate
and the density of states shown in Figs.~\ref{Fig:DOSU6} 
to~\ref{Fig:DOSU4c5} have a much better resolution than 
those calculated with other methods such as the
FE-DMFT(ED)~\cite{mottinsu},
DMFT(NRG)~\cite{NRG} or DMFT(DMRG)~\cite{RozenbergHallberg};
DMFT(QMC) calculations~\cite{Bluemer} have not provided estimates for
these quantities.
In particular, our investigation demonstrates
the presence of a
sharp feature just above the gap in the density of states of the 
insulator and, thus, provides the first clear evidence for a
splitting of the quasi-particle peak at the metal-insulator
transition.
As an accurate description of the low-energy excitations
is necessary close to the critical coupling $U_{\rm c}$, 
it is not surprising
that the insulating phase extends to slightly 
weaker couplings than predicted in previous 
works~\cite{NRG,Bluemer,RozenbergHallberg}.
The impurity solver used in previous DMFT investigations do not have the
accuracy of the DDMRG method combined with the `fixed-energy' 
dynamical mean-field theory. Therefore, they could not resolve 
the small gap $\Delta(U) \leq 0.260$ for $U \leq 4.7$ and did
not find a stable insulating phase below that coupling.

\section{Summary and conclusions}
\label{summary}

We have investigated the paramagnetic insulating
ground state of the Hubbard model on a Bethe
lattice in the limit of infinite coordination number. 
In this limit, the problem can be treated within the 
dynamical mean-field theory (DMFT), i.e., it can be mapped onto
a system made of a single impurity with Hubbard interaction
and hybridizations to a bath.
The system properties have to be determined self-consistently
from the required equivalence between the single-particle Green function
and the hybridization function.
In this work, we have used the fixed-energy approach to the DMFT 
(FE-DMFT)~\cite{mottinsu}.
The FE-DMFT provides stable solutions of the
DMFT self-consistency problem
and a systematic convergence of the results with
increasing system size.

As `impurity solver' for the single-particle density of states
we have used the dynamical density-matrix
renormalization group (DDMRG)~\cite{EricSatoshi}.
Our results from FE-DMFT(DDMRG) 
for the ground state
agree with perturbation theory in $1/U$, where the latter is applicable,
and with quantum Monte Carlo (QMC) simulations, where the DMFT(QMC)
approach finds a stable insulating phase.
With DDMRG we have used up to 160 sites to represent
the self-consistent DMFT bath 
as compared to $n_s=15$ with exact diagonalization (ED).
These larger systems provide an enhanced resolution
which is necessary to reveal structures in the
density of states near the gap.
These structures emerge when the Mott gap becomes small,
i.e., when the coupling $U$ approaches a critical
value $U_{\rm c}$ below which there is no insulating phase
($\equiv U_{{\rm c},1}$ in a scenario with coexisting metallic
and insulating phases).
Our FE-DMFT(DDMRG) study gives~$U_{\rm c}=4.45\pm 0.05$
for the critical interaction strength where the gap closes,
in very good agreement with our previous FE-DMFT(ED)
study~\cite{mottinsu}, $U_{\rm c}=4.43\pm 0.05$.

In contrast to the results of a recent DMFT(DMRG)
work~\cite{RozenbergHallberg},
our results are not dominated by finite-size effects. 
At $U=6$, for example,
the density of states in~\cite{RozenbergHallberg}
displays a series of individual peaks instead of the smooth
Hubbard bands found in our approach and in perturbation theory.
Preliminary results for the metallic Fermi-liquid
phase just below~$U_{\rm c}$
suggest that the narrow quasi-particle resonance simply splits 
at~$U_{\rm c}$. 
Narrow shoulders which can be seen in the insulator
density of states for $U=4.5$ in Fig.~\ref{Fig:DOSU4c5}
are the remainders of the quasi-particle resonance. 
The shoulders quickly loose weight as the upper and lower 
Hubbard bands separate from each other with 
increasing interaction strength~$U$.

The method presented here can also be applied to the metallic 
phase, as done in Ref.~\cite{mottmetal} for the weak-coupling limit.
It is more difficult to resolve sharp quasi-particle peaks
with DDMRG~\cite{EricSatoshi} than with, e.g., the
numerical renormalization group.
However, as shown in this work, 
our method offers the unique advantage that we can resolve
sharp structures in the vicinity of the Hubbard band onsets.
This is very important to describe the Mott insulating phase accurately
and to determine the parameter region where it exists.
Therefore, we are confident that our FE-DMFT(DDMRG) will provide
deeper insight into the Mott--Hubbard metal-to-insulator transition.

\ack
We thank N. Bl\"umer for helpful discussions. This work was supported by
the center {\sl Optodynamik\/} of the Philipps-Universit\"at Marburg.

\appendix

\section{Sum rule}
\label{appA}

At self-consistency we have from equations~(\ref{Dforlateruse}),
(\ref{Mndef}), (\ref{sumrule1}), and~(\ref{selfcons})
\begin{eqnarray}
\fl 
M_1(U)-\frac{U}{2}&=& \sum_{\sigma} \int_{\mu_{\rm LHB}^-}^{\mu_{\rm LHB}^+}
\diff \omega \, \omega \, D_{\rm LHB}(\omega) \nonumber \\
\fl
&=& \sum_{\sigma} \int_{\mu_{\rm LHB}^-}^{\mu_{\rm LHB}^+}
\diff \omega \omega \left \langle \hat{d}_{\sigma}^+
\delta\left(\omega+\hat{H}_{\rm SIAM}-E_0^{\rm SIAM}\right) \hat{d}_{\sigma}
\right \rangle_{\rm SIAM} \label{basiseq}\\
\fl
&=& \sum_{\sigma} \langle \hat{d}_{\sigma}^+
\left[ \hat{d}_{\sigma}, \hat{H}_{\rm SIAM}\right]_{-}
\rangle_{\rm SIAM} \; , \nonumber
\end{eqnarray}
where $E_0^{\rm SIAM}$ is the ground-state energy of the single-impurity
Anderson model. The fact that the first moments are identical 
at self-consistency also
proves that the average double occupancy of the Hubbard model~$d(U)$
is identical to the average double occupancy of the interacting site
in the impurity model. Therefore, we do not distinguish 
between~$d(U)$ and ~$d_{\rm SIAM}(U)$.

We carry out the commutators in~(\ref{basiseq}) using the 
Hamiltonian~(\ref{SIAMns}) and obtain
\begin{equation}
M_1(U)= 2 U d(U) + \sum_{\ell,\sigma} V_{\ell}
\langle \hat{d}_{\sigma}^+\hat{\psi}_{\ell,\sigma} \rangle_{\rm SIAM} \; .
\label{M1siam}
\end{equation}
The ground-state expectation value on the right-hand-side
of this equation is readily calculated in DMRG.
Therefore, combining~(\ref{sumrule2}), (\ref{dbar}) and~(\ref{M1siam})
provides the ground-state energy density of the Hubbard model
in terms of the single-impurity results as
\begin{equation}
\frac{E_0(U)}{L} = U d(U) - \frac{U}{4} 
+ \sum_{\ell,\sigma} V_{\ell}
\langle \hat{d}_{\sigma}^+\hat{\psi}_{\ell,\sigma} \rangle_{\rm SIAM} \; ,
\end{equation}
which is equivalent to~(\ref{energy}).

The static ground-state expectation value in~(\ref{basiseq})
can be obtained from the corresponding Green function.
For completeness we define the time-ordered Green functions
\begin{eqnarray}
G_{\ell\ell;\sigma}(t) &=& -\I \langle \hat{\cal T} [ 
\hat{\psi}_{\ell,\sigma}(t)\hat{\psi}_{\ell,\sigma}^+] \rangle_{\rm SIAM}
 \; , \nonumber \\
G_{\ell d;\sigma}(t) &=& -\I \langle \hat{\cal T} [ 
\hat{\psi}_{\ell,\sigma}(t)\hat{d}_{\sigma}^+] \rangle_{\rm SIAM}
 \; , \label{missingGF}\\
G_{d\ell;\sigma}(t) &=& -\I \langle \hat{\cal T} [ 
\hat{d}_{\sigma}(t)\hat{\psi}_{\ell,\sigma}^+] \rangle_{\rm SIAM}
 \; . \nonumber
\end{eqnarray}
With the help of the equation of motion it is not difficult
to show that their Fourier transforms 
obey ($\omega\equiv \omega+\I0^+\sgn \, \omega$)
\begin{eqnarray}
G_{\ell\ell;\sigma}(\omega) &=& \frac{1}{\omega-\epsilon_{\ell}} +
\frac{V_{\ell}^2}{(\omega-\epsilon_{\ell})^2} G_{dd;\sigma}(\omega) 
\nonumber \; ,\\
G_{\ell d;\sigma}(\omega) &=& \frac{V_{\ell}}{\omega-\epsilon_{\ell}} 
G_{dd;\sigma}(\omega) 
= G_{d\ell;\sigma}(\omega) \; . \label{missingGFomega} 
\end{eqnarray}
Then, the first moment in~(\ref{M1siam}) becomes ($\nu=0^+$)
\begin{equation}
M_1 = 2 Ud(U) + \sum_{\ell,\sigma} V_{\ell} \int_{-\infty}^{\infty}
\frac{\diff\omega}{2\pi \I} e^{\I\omega\nu} 
\frac{V_{\ell}}{\omega-\epsilon_{\ell}} G_{dd;\sigma}(\omega) \; .
\end{equation}
At self-consistency we have~(\ref{selfcons}) which implies
\begin{equation}
\sum_{\sigma} H(\omega) = \sum_{\ell,\sigma} 
\frac{V_{\ell}^2}{\omega-\epsilon_{\ell}}
= G(\omega) \; . \label{HandGlink}
\end{equation}
Therefore, we arrive at the important relation
\begin{equation}
\int_{-\infty}^{\infty} \frac{\diff\omega}{2\pi \I} e^{\I\omega\nu} 
\left[ G(\omega) \right]^2 = 2\left(M_1(U)-2Ud(U)\right) \; .
\label{integralandM1}
\end{equation}
It relates the ground-state energy of the single-impurity model
back to the ground-state energy of the Hubbard model.
To show this we start from
\begin{eqnarray}
E_0^{\rm SIAM}(U) &=& -\frac{U}{4} + U d(U) +
\sum_{\ell,\sigma} V_{\ell} \langle \hat{\psi}_{\ell,\sigma}^+\hat{d}_{\sigma}
+ \hat{\psi}_{\ell,\sigma}\hat{d}_{\sigma}^+\rangle_{\rm SIAM} \nonumber \\
&& + \sum_{\ell,\sigma} \epsilon_{\ell} 
\langle \hat{\psi}_{\ell,\sigma}^+\hat{\psi}_{\ell,\sigma}
\rangle_{\rm SIAM} \; .
\label{defE0SIAM}
\end{eqnarray}
With the help of~(\ref{missingGFomega}) and~(\ref{HandGlink})
we find for the second term in~(\ref{defE0SIAM})
\begin{equation}
\sum_{\ell,\sigma} V_{\ell} \langle \hat{\psi}_{\ell,\sigma}^+\hat{d}_{\sigma}
+ \hat{\psi}_{\ell,\sigma}\hat{d}_{\sigma}^+\rangle_{\rm SIAM}
= 2\left(M_1(U)-2Ud(U)\right) \; .
\label{result2}
\end{equation}
For the third term we use 
\begin{equation}
\sum_{\ell,\sigma} \frac{V_{\ell}^2 
\epsilon_{\ell}}{(\omega-\epsilon_{\ell})^2}
= -H(\omega) -\omega \frac{\partial H(\omega)}{\partial \omega}
\end{equation}
to obtain
\begin{eqnarray}
\sum_{\ell,\sigma} \epsilon_{\ell} 
\langle \hat{\psi}_{\ell,\sigma}^+\hat{\psi}_{\ell,\sigma}
\rangle_{\rm SIAM} &=&
\sum_{\ell,\sigma} \epsilon_{\ell} \Theta(-\epsilon_{\ell}) 
\label{result3} \\
&& -\frac{1}{2} \int_{-\infty}^{\infty} 
\frac{\diff\omega}{2\pi \I} e^{\I\omega\nu} G(\omega) 
\left[ G(\omega) +\omega \frac{\partial G(\omega)}{\partial \omega}
\right] \nonumber \\
&= & \sum_{\ell,\sigma} \epsilon_{\ell} \Theta(-\epsilon_{\ell}) 
-\frac{1}{2} \left(M_1(U)-2Ud(U)\right) \; .
\nonumber            
\end{eqnarray}
Summing the contributions from~(\ref{result2}) and~(\ref{result3})
in~(\ref{defE0SIAM}) gives the final result
\begin{equation}
E_0^{\rm SIAM}(U) = \sum_{\ell,\sigma} \epsilon_{\ell} 
\Theta(-\epsilon_{\ell}) 
-2 U d(U) + \frac{3}{2} M_1(U) -\frac{U}{4} 
\; ,
\end{equation}
where $d(U)$ is the average double occupancy~(\ref{dbar}),
$M_1(U)$ is the first moment~(\ref{sumrule2}),
and $\Theta(x)$ is the step function.
This equation expresses the fact that the impurity provides corrections
of order unity to the extensive ground-state energy. 
Therefore, for our equidistant energy levels~$\epsilon_{\ell}$ we find
equation~(\ref{eqfromappA}).

\end{document}